\documentclass{PoS}
\usepackage{slashed}

\title{Constraints on neutrino millicharge and charge radius from neutrino-atom scattering}

\ShortTitle{Constraints on neutrino millicharge and charge radius}

\author{Matteo Cadeddu\\
        Universit\`{a} degli Studi di Cagliari and Istituto Nazionale di Fisica Nucleare (INFN), Sezione di Cagliari, Complesso Universitario di Monserrato--S.P.
per Sestu Km 0.700, 09042 Monserrato (Cagliari), Italy\\
        E-mail: \email{matteo.cadeddu@ca.infn.it}}

\author{Carlo Giunti\\
        Istituto Nazionale di Fisica Nucleare (INFN), Sezione di Torino, Via P. Giuria 1, I-10125 Torino, Italy\\
        E-mail: \email{carlo.giunti@to.infn.it}}

\author{Konstantin A. Kouzakov\\
        Department of Nuclear Physics and Quantum Theory of Collisions, Faculty of Physics, Lomonosov Moscow State University, Moscow 119991, Russia\\
        E-mail: \email{kouzakov@gmail.com}}

\author{Yufeng Li\\
        Institute of High Energy Physics, Chinese Academy of Sciences, Beijing 100049, China
and School of Physical Sciences, University of Chinese Academy of Sciences,
Beijing 100049, China\\
        E-mail: \email{liyufeng@ihep.ac.cn}}

\author{\speaker{Alexander I. Studenikin}\\
        Department of Theoretical Physics, Faculty of Physics, Lomonosov Moscow State University, Moscow 119991, Russia\\
        Joint Institute for Nuclear Research, Dubna 141980, Moscow Region, Russia\\
        E-mail: \email{studenik@srd.sinp.msu.ru}}

\author{Yiyu Zhang\\
        Institute of High Energy Physics, Chinese Academy of Sciences, Beijing 100049, China
and School of Physical Sciences, University of Chinese Academy of Sciences,
Beijing 100049, China\\
        E-mail: \email{zhangyiyu@ihep.ac.cn}}

\abstract{We consider possible effects of neutrino electric charge (millicharge) and charge radius on the neutrino-atom
interaction processes such as (i) atomic ionization by neutrino impact and (ii) coherent elastic neutrino-nucleus
scattering. The bounds on the neutrino millicharge and charge radius that follow from, respectively,
the GEMMA and COHERENT experiments are presented and discussed.}

\FullConference{
European Physical Society Conference on High Energy Physics - EPS-HEP2019 -\\
			10-17 July, 2019\\
			Ghent, Belgium}

\begin{document}

\section{Introduction}
%
The development of our knowledge about neutrino masses and mixing provides a basis for exploring neutrino properties and interactions beyond the standard model (BSM).
In this respect, the study of nonvanishing electromagnetic characteristics of massive neutrinos~\cite{bib:Giunti,andp16}
can help to constrain the existing BSM theories and/or to hint at new physics.
The effects of neutrino electromagnetic properties can be searched in astrophysical environments, where neutrinos propagate in strong magnetic fields and dense matter, and in laboratory measurements of neutrinos from various sources. In the latter case, a very sensitive method is provided by the direct measurement of low-energy elastic neutrino scattering on atomic electrons and nuclei in a detector.
In this contribution, we present our bounds on the neutrino millicharge and charge radii that have been derived from the data of the GEMMA~\cite{GEMMA:2012} and COHERENT~\cite{COHERENT:2017} scattering experiments, respectively, and included in the Particle Data Group's Review of Particle Physics~\cite{PDG2016}.

\section{Electromagnetic properties of massive neutrinos}
%
There are at least three massive neutrino fields $\nu_{i}$ with respective masses
$m_{i}$ ($i=1,2,3$), which are mixed with the three active flavor neutrinos $\nu_{e}$,
$\nu_{\mu}$, $\nu_{\tau}$.
Therefore, the neutrino effective
electromagnetic vertex, which in momentum-space representation depends
only on the four-momentum $q=p_i-p_f$ transferred to the photon, can be presented as follows~\cite{bib:Giunti,andp16}:
\begin{equation}
\Lambda_{\mu}(q) = \left( \gamma_{\mu} - q_{\mu}
\slashed{q}/q^{2} \right) \left[ f_{Q}(q^{2}) + f_{A}(q^{2})
q^{2} \gamma_{5} \right]  - i \sigma_{\mu\nu} q^{\nu} \left[ f_{M}(q^{2}) +
i f_{E}(q^{2}) \gamma_{5} \right]. \label{C043}
\end{equation}
Here $\Lambda_{\mu}(q)$ is a $3{\times}3$ matrix in the space
of massive neutrinos expressed in terms of the four Hermitian
$3{\times}3$ matrices of form factors
$f_{Q} = f_{Q}^{\dagger}$, $f_{M} = f_{M}^{\dagger}$,
$f_{E} = f_{E}^{\dagger}$, and $f_{A} = f_{A}^{\dagger}$,
%
%
where $Q,M,E,A$ refer respectively to the real charge, magnetic, electric, and anapole neutrino form factors.

For the coupling with a real photon in vacuum ($q^{2}=0$) one has
$f_{Q}^{fi}(0) = e_{fi}$, $f_{M}^{fi}(0) = \mu_{fi}$, $f_{E}^{fi}(0) = \epsilon_{fi}$, and $f_{A}^{fi}(0) = a_{fi}$,
where $e_{fi}$, $\mu_{fi}$, $\epsilon_{fi}$ and $a_{fi}$ are,
respectively, the neutrino charge, magnetic moment, electric
moment and anapole moment of diagonal ($f=i$) and transition
($f{\neq}i$) types. Even if the electric charge of a neutrino is zero,
$f_Q(q^2)$ can still contain nontrivial
information about neutrino electrostatic properties, namely the neutrino charge radius.
The mean charge radius (in fact, it is the squared
charge radius) of an electrically neutral neutrino is
given by
\begin{equation}
\langle r^2_\nu\rangle=\frac{1}{6}\,\left.\frac{df_Q(q^2)}{dq^2}\right|_{q^2=0}.
\end{equation}
%

%
\section{Elastic neutrino-electron scattering}
Here we consider the process $\nu+e^-\to e^-+\nu$ where an ultrarelativistic neutrino 
with energy $E_\nu$ 
elastically scatters on an electron in a detector at energy transfer $T$.
In the scattering experiments the observables are the kinetic energy $T_e$ of the recoil electron and/or its solid angle $\Omega_e$. From the energy-momentum conservation one gets
\begin{equation}
T_e=T, \qquad \cos\theta_e=\left(1+\frac{m_e}{E_\nu}\right)\sqrt{\frac{T}{T+2m_e}},
\end{equation}
where $\theta_e$ is the angle of the recoil electron with respect to the neutrino beam. 
The cross section, which is differential with respect to the electron kinetic energy $T_e$, can be presented in the form of a sum of helicity-conserving ($w,Q$) and helicity-flipping ($\mu$) components~\cite{Kouzakov:2017prd}:
\begin{equation}
\label{cr_sec_FE_1}
\frac{d\sigma}{dT_e}=\frac{d\sigma_{(w,Q)}}{dT_e}+\frac{d\sigma_{(\mu)}}{dT_e},
\end{equation}
where $d\sigma_{(w,Q)}/dT_e$ is the electroweak cross section modified by the effect of the neutrino millicharge, charge radius and anapole moment, and $d\sigma_{(\mu)}/dT_e$ is the magnetic cross section due to the neutrino dipole magnetic and electric moments.

At small $T_e$ values the contributions to the recoil-electron spectrum due to the weak, millicharge, and magnetic scattering channels exhibit qualitatively different $T_e$ dependencies, namely
\begin{equation}
\mathcal{N}_{e^-}^{(w,Q)}(T_e)\propto\left\{
\begin{array}{cr} \displaystyle {\rm const} & (e_\nu=0),\\
\displaystyle\frac{2\pi\alpha^2}{m_eT_e^2}\left(\frac{e_\nu}{e_0}\right)^2 & (e_\nu\neq0),
\end{array}\right. \qquad {\rm and} \qquad \mathcal{N}_{e^-}^{(\mu)}(T_e)\propto\frac{\pi\alpha^2}{m_e^2T_e}\left(\frac{\mu_\nu}{\mu_B}\right)^2,
\end{equation}
where $\alpha$ is the fine structure constant, $e_\nu$ and $\mu_\nu$ are the neutrino (effective) millicharge and magnetic moment, and $e_0$ and $\mu_B$ are an elementary electric charge and a Bohr magneton, respectively. For the ratio $\mathcal{R}$ of the millicharge and magnetic-moment contributions to the recoil-electron energy spectrum one thus has
\begin{equation}
\label{ratio}
\mathcal{R}=\frac{\mathcal{N}_{e^-}^{(Q)}(T_e)}{\mathcal{N}_{e^-}^{(\mu)}(T_e)}=\frac{2m_e}{T_e}\,\frac{({e_\nu}/{e_0})^2}{({\mu_\nu}/{\mu_B})^2}.
\end{equation}
In case there are no observable deviations from the weak contribution to the electron spectrum it is possible to get the upper bound for the neutrino millicharge demanding that a possible effect due to $e_\nu$ does
not exceed one due to the neutrino (anomalous) magnetic moment $\mu_\nu$. This implies that $\mathcal{R}<1$ and from the relation~(\ref{ratio}), 
%
using the GEMMA data~\cite{GEMMA:2012}, namely the detector energy threshold $\sim2.8$ keV and the $\mu_\nu$ bound $\mu_\nu<2.9\times10^{-11}\mu_B$, one obtains the following upper limit on the neutrino millicharge~\cite{Studenikin:2014epl}:
$$
|e_\nu|<1.5\times10^{-12}e_0.
$$
The $e_\nu$ range that expected to be probed in a few years with the
GEMMA-II experiment (an effective threshold of 1.5 keV and the $\mu_\nu$ sensitivity at the level of
$1\times10^{-11}\mu_B$) is $|e_\nu|<3.7\times10^{-13}e_0$.

\section{Coherent elastic neutrino-nucleus scattering}
Here we consider the process $\nu_\ell+a(Z,N)\to{a}(Z,N)+\nu_{\ell'=e,\mu,\tau}$ where an utrarelativistic neutrino with energy $E_\nu$ elastically scatters on an atomic nucleus, having $Z$ protons and $N$ neutrons, in a detector at energy-momentum transfer $q=(T,\vec{q})$. For a spin-zero nucleus and $T_{a}\ll E_\nu$, where $T_{a}=T$ is the nuclear recoil kinetic energy, the differential cross section due to the weak and charge-radius scattering channels is given by~\cite{Kouzakov:2017prd,Caddedu:2018prd}
\begin{equation}
\label{coherent_cross_section}
\frac{d\sigma_{(w,r_\nu)}}{dT_{a}}\simeq\frac{G_F^2M_a}{\pi}\left(1-\frac{M_aT_{a}}{2E_\nu^2}\right)
\left\{\left[\left(g_V^p-\delta_{\ell\ell}\right)F_Z(|\vec{q}|^2)+g_V^nF_N(|\vec{q}|^2)\right]^2+
F_Z^2(|\vec{q}|^2)\sum_{\ell'\neq\ell}|\delta_{\ell\ell'}|^2 \right\},
\end{equation}
where $M_a$ is the nuclear mass, $g_V^p=1/2-2\sin^2\theta_W$ and $g_V^n=-1/2$ (the neglected radiative corrections are too small to affect the results). $F_{Z,N}(|\vec{q}|^2)$, such that $F_Z(0)=Z$ and $F_N(0)=N$, are the nuclear form factors, which are the Fourier transforms of the corresponding nucleon density distribution in the nucleus and describe the loss of coherence for $|\vec{q}|R\gtrsim1$, where $R$ is the nuclear radius. The effect of the neutrino charge radii is accounted for through
$$
\delta_{\ell\ell'}=\frac{2}{3}\,m_W^2\sin^2\theta_W\langle r_{\nu_{\ell\ell'}}^2\rangle, \qquad {\rm with} \qquad
\langle r_{\nu_{\ell\ell'}}^2\rangle=\sum_{i,j}U_{\ell i}^*U_{\ell' j}\langle r_{\nu_{ij}}^2\rangle,
$$
where $U$ is the neutrino mixing matrix. The diagonal $(\ell=\ell')$ charge radii are already predicted in the standard model~\cite{Bernabeu}:
\begin{equation}
\label{r_nu_SM}
\langle r_{\nu_e}^2\rangle_{\rm SM}=-0.83\times10^{-32}~{\rm cm}^2, \quad
\langle r_{\nu_\mu}^2\rangle_{\rm SM}=-0.48\times10^{-32}~{\rm cm}^2, \quad
\langle r_{\nu_\tau}^2\rangle_{\rm SM}=-0.30\times10^{-32}~{\rm cm}^2.
\end{equation}
However, the transition $(\ell\neq\ell')$ charge radii are essentially the BSM quantities.

The results of our fit of the time-dependent COHERENT data~\cite{COHERENT:2017} are presented in Ref.~\cite{Caddedu:2018prd}.
In addition to the customary, diagonal charge radii, from the COHERENT data we have obtained for the first time limits on the neutrino transition charge radii~\cite{Caddedu:2018prd}:
$$
\left(|\langle r_{\nu_{e\mu}}^2\rangle|,|\langle r_{\nu_{e\tau}}^2\rangle|,|\langle r_{\nu_{\mu\tau}}^2\rangle|,\right)
< (22,38,27)\times10^{-32}~{\rm cm}^2,
$$
at 90\% CL, marginalizing over reliable allowed intervals of the rms radii $R_n(^{133}{\rm Cs})$ and $R_n(^{127}{\rm I})$. This is an interesting information on the BSM physics which can generate the neutrino transition charge radii~\cite{Novales-Sanchez:2008prd}.


\begin{thebibliography}{99}
%
\bibitem{bib:Giunti} C. Giunti and A. Studenkin, \emph{
Rev. Mod. Phys.} \textbf{87} (2015) 531.
%
\bibitem{andp16} C. Giunti, K. A. Kouzakov, Y.-F. Li, A. V. Lokhov, A. I. Studenikin and S. Zhou, \emph{
Ann. Phys. (Berlin)} \textbf{528} (2016) 198.
%
\bibitem{GEMMA:2012} A. Beda \emph{et al.} (GEMMA Collaboration), \emph{Adv. High Energy Phys.} \textbf{2012} (2012) 350150.
%
\bibitem{COHERENT:2017} D. Akimov \emph{et al.} (COHERENT Collaboration), \emph{Science} \textbf{357} (2017) 1123.
%
\bibitem{PDG2016} C. Patrignani \emph{et al.} (Particle Data Group), \emph{Chin. Phys. C} \textbf{40} (2016) 100001; M. Tanabashi \emph{et al.} (Particle Data Group), \emph{Phys. Rev. D} \textbf{98} (2018) 030001 and 2019 update.
%
\bibitem{Kouzakov:2017prd} K. A. Kouzakov and A. I. Studenikin, \emph{
Phys. Rev. D} \textbf{95} (2017) 055013.
%
\bibitem{Studenikin:2014epl} A. Studenikin, \emph{Europhys. Lett.} \textbf{107} (2014) 21001.
%
\bibitem{Caddedu:2018prd} M. Caddedu, C. Giunti, K. A. Kouzakov, Y.-F. Li, A. I. Studenikin and Y. Y. Zhang, \emph{
Phys. Rev. D} \textbf{98} (2018) 113010.
%
\bibitem{Bernabeu} J. Bernabeu, L. G. Cabral-Rosetti, J. Papavassiliou and J. Vidal, \emph{
Phys. Rev. D} \textbf{62} (2000) 113012;
J. Bernabeu, J. Papavassiliou and J. Vidal, \emph{Phys. Rev. Lett.} \textbf{89} (2002) 101802; \emph{Nucl. Phys.} \textbf{B680} (2004) 450.
%
%
%
\bibitem{Novales-Sanchez:2008prd} H. Novales-Sanchez, A. Rosado, V. Santiago-Olan and J. Toscano, \emph{Phys. Rev. D} \textbf{78} (2008) 073014.
%
\end{thebibliography}
\end{document}